# Efficient reconciliation protocol for discrete-variable quantum key distribution

David Elkouss*, Anthony Leverrier*, Romain Alléaume* and Joseph J. Boutros†
*Institut TELECOM, TELECOM ParisTech & LTCI, UMR CNRS 5141,
37/39, Rue Dareau, 75014 Paris, France.
Email: {elkouss, leverrier, alleaume}@telecom-paristech.fr
†Coding, Communications, and Information Theory, Texas A&M University at Qatar,
Education City, PO Box 23874, Doha, Qatar.
Email: boutros@tamu.edu

*Abstract*—Reconciliation is an essential part of any secret-key agreement protocol and hence of a Quantum Key Distribution (QKD) protocol, where two legitimate parties are given correlated data and want to agree on a common string in the presence of an adversary, while revealing a minimum amount of information.

In this paper, we show that for discrete-variable QKD protocols, this problem can be advantageously solved with Low Density Parity Check (LDPC) codes optimized for the BSC. In particular, we demonstrate that our method leads to a significant improvement of the achievable secret key rate, with respect to earlier interactive reconciliation methods used in QKD.

## I. INTRODUCTION

In a QKD protocol [1], two legitimate parties, Alice and Bob, aim at sharing an information theoretic secret key, even in the presence of an eavesdropper Eve. In the quantum part of such a protocol, Alice and Bob exchange quantum signals, e.g. single photons, which carry classical information. For instance, Alice encodes a classical bit onto the polarization or the phase of a photon and sends this photon to Bob who measures it. After repeating this step $n$ times, Alice and Bob share two $n-$bit strings, $X$ and $Y$. Eve has access to a random variable $Z$, possibly correlated to $X$ and $Y$.

In any realistic implementation of a QKD protocol, $X$ and $Y$ suffer discrepancies mainly due to losses in the channel and noise in Bob's detectors but which are conservatively attributed to the action of an eavesdropper. Therefore, any QKD protocol must include a classical post-processing step in order to extract a secret key from the correlated strings $X$ and $Y$. This is done thanks to classical communication over a noiseless, authenticated but otherwise insecure channel. This secret key agreement is itself split in two parts. First, Alice and Bob correct the errors between their strings: this is the so-called *reconciliation* phase which concerns us here. Then, in the *privacy amplification* phase [2], Alice and Bob apply a randomly chosen compression function to their mutual string. If the compression function is well chosen, the result is uncorrelated with $Z$ and constitutes a secret key.

The theoretical *secret capacity* $K_{\text{th}}$ is given by:

$$K_{\text{th}} = H(X|Z) - H(X|Y). \quad (1)$$

The precise definition of $H(X|Z)$ depends on the type of attack considered, whereas $H(X|Y)$ represents the conditional (Shannon) entropy of $X$ given $Y$.

This secret capacity is actually theoretical and is achieved only in the case of a perfect reconciliation scheme. In particular, the term $H(X|Y)$ corresponds to the minimum amount of classical information that Alice needs to send to Bob to help him correct his string $Y$. In a realistic implementation, the actual secret key rate $K_{\text{real}}$ is given by:

$$K_{\text{real}} = H(X|Z) - fH(X|Y), \quad (2)$$

where $f$ is a parameter greater than 1 that characterizes the reconciliation efficiency.

The main effect of an imperfect reconciliation is clearly a reduction of the secret key rate, which in turn, limits the range of the QKD protocol. This is the reason why the reconciliation should be as efficient as possible. However, one should keep in mind two other important factors when evaluating a reconciliation scheme: its complexity and its rapidity. This last criterium is especially relevant in the case of highly interactive schemes where latency can become an issue.

In most QKD protocols, the information is encoded on binary variables. This is the case we will consider here. Errors are usually uncorrelated and symmetric. For this reason, $X$ and $Y$ can be seen, respectively, as the input and the output of a binary symmetric channel (BSC). In a typical implementation of a QKD protocol, Alice and Bob have access to the channel characteristics. In particular, the crossover probability $p$ of the BSC is supposed known.

To fix ideas, let us consider the most emblematic QKD protocol: BB84 [3]. For this protocol, the different conditional entropies can be easily expressed as a function of $p$: $H(X|Z) = 1 - h(p)$ and $H(X|Y) = h(p)$ where $h(p) = -p \log_2 p - (1-p) \log_2(1-p)$ is the binary entropy function. This leads to:

$$K_{\text{real}} = 1 - (1 + f(p))h(p). \quad (3)$$

In the following, we will use this expression to compare the efficiencies of different reconciliation schemes. One should note that even with a perfect reconciliation scheme, the maximum bit error rate admissible to distribute a secret is 11%. Typical implementations have an error rate between 3 and 10 %. This is this range of parameter that interests us here.

The rest of the paper is organised as follows: in section II, we review the Cascade protocol which is currently the solution adopted in most implementations. In section III, we present an optimization technique of LDPC codes for the BSC. The respective performances of Cascade and our LDPC codes are then discussed in section IV.

## II. PREVIOUS WORK

### A. The Cascade protocol

Cascade was proposed by Brassard and Salvail in their seminal paper "Secret key reconciliation by public discussion" [4] as an alternative to error correcting codes because at the time their complexity was too high to be used in practice [5]. Cascade takes benefit from the interaction between Alice and Bob over an authenticated public channel to simplify the problem of reconciliation. It can be described by a very compact and elegant algorithm.

The Cascade protocol is run iteratively a given number of passes, this number being determined as a function of the estimated probability of error. This error estimation is conducted prior to beginning the protocol, on a statistically significant random sample of Alice and Bob's data. In each pass $i$, Alice and Bob agree on a random permutation $\sigma_i$ which they apply to their strings: $X^i = \sigma_i(X)$ and $Y^i = \sigma_i(Y)$. Then they divide their permutated strings into blocks of $k_i$ bits. After each pass the block size will be doubled: $k_i = 2k_{i-1}$. The value of the initial block size $k_1$ is a critical parameter. An empirical result in [5] indicates that an optimal value is $k_1 \approx 0.73/e$, $e$ being the estimated error probability.

For each block $j$ Alice sends its parity $x_j$ to Bob while Bob computes $y_j$, the parity of its block, sends it to Alice, and compares it with $x_j$. If $y_j \neq x_j$ Alice and Bob perform a binary search to find and correct an error in position $p$. The binary search consists in splitting the block $j$ into two halves, and then calculate and exchange the parity of one half. If both parities do not agree, Alice and Bob continue the binary search with the same half, if they agree they continue with the other half.

The position $p$ where an error has been found belonged to different blocks in the preceding passes. Let $C$ be the set of such blocks with an odd number of errors. Alice and Bob can now choose the smallest block in $C$ and perform a binary search to find and correct another error. This new error will imply adding or removing blocks from $C$. This process is performed until $C$ is emptied.

It should be noted that Cascade is highly interactive even when carefully implemented. Since many exchanges between Alice and Bob are required to reconcile a string, the time overhead for these communications can severely limit the achievable key generation rate. This could for instance be the case in free space QKD implemented between a satellite and a base station and even more when the communication between Alice and Bob is performed over a network connection with a high latency.

Despite this limitation, Cascade is certainly the most widely used reconciliation protocol in practical discrete variables QKD setups. One of its interests is its relative simplicity and the fact that it performs reasonably well in terms of efficiency. As we shall see, most of the alternative solutions developed after Cascade have focused on reducing the level of interactivity, usually at the expense of reconciliation efficiency. This is the reason why we have used Cascade as the essential element of comparison with the solution we have designed, that has the double advantage of being non-interactive and of performing better that Cascade in terms of efficiency over a wide parameter range.

### B. Other work on information reconciliation protocols

Many variations around the principle of interactive reconciliation used in Cascade have been proposed, in order to limit the interactivity. Relevant work on the optimization of the block lengths has been done in [6], and allows to limit the number of rounds in the regime of very low error rate. Among the most notable works, we can also cite the Winnow protocol[7]. Like Cascade, Winnow splits the binary strings to be reconciled into blocks but instead of correcting errors by iterative binary search, the error correction is based on the Hamming code. Winnow's interest lies in the reduction of the amount of required communication to three messages per iteration [8]. Winnow is thus significantly faster that Cascade but unfortunately, its efficiency is lower for error rates below 10 %, i.e. in the parameter range useful for practical QKD. Another interesting development has been conducted by Liu[9] who has proposed a protocol that optimizes the information exchanged per corrected bit. Liu's protocol is in essence very similar to Cascade. Its objective is to minimize the information sent on the public channel to correct one error during a pass and leads to better efficiency. This protocol however remains highly interactive.

Some QKD protocols provide Alice and Bob with correlated continuous random variables and specific work on key reconciliation has been conducted in this context, beginning with the work on Sliced Error Correction [10] used to convert continuous variables into binary strings. It is also mainly in the context of continuous variables that modern coding techniques have been used within information reconciliation protocols: turbo codes in [8] and LDPC codes in [11], [12].

In contrast with continous-variable information reconciliation, not much has been done to adapt modern coding techniques to the discrete case. Forward error correction has the advantage of being very well known and even attaining the Shannon limit for some channels. Also, and of great importance for QKD, it requires a single message, namely the syndrome of $X$ for the code being used, to correct the discrepancies. Relevant references are BBN Niagara [13] and the work for free space QKD by Duligall et al. [14] both of which use LDPC codes. However [13] provides a single point comparing the performance of LDPC codes and Cascade, showing a major decrease of the communication overhead but a slightly decrease in the efficiency while [14] does not provide any information on the results of their use of LDPC codes.

## III. OPTIMIZATION OF LDPC CODES FOR THE BSC

LDPC codes also known as Gallager codes are linear codes that have a sparse parity check matrix, that is with relatively

TABLE I
THRESHOLDS AND DEGREE DISTRIBUTIONS FOUND FOR A REPRESENTATIVE SET OF RATES

| Code rate | Threshold | $\lambda(x)$ & $\rho(x)$ |
|---|---|---|
| 0.90 | 0.0109 | $\lambda(x) = 0.07689x + 0.28096x^2 + 0.08933x^4 + 0.19620x^8 + 0.30631x^{11} + 0.05031x^{20}$<br>$\rho(x) = 0.95025x^{49} + 0.04975x^{50}$ |
| 0.85 | 0.0199 | $\lambda(x) = 0.04528x + 0.20537x^2 + 0.05878x^3 + 0.094274x^4 + 0.08454x^5 + 0.01176x^6 + 0.05137x^8 + 0.50015x^{20}$<br>$\rho(x) = 0.54204x^{40} + 0.45795x^{41}$ |
| 0.80 | 0.0298 | $\lambda(x) = 0.09420x + 0.18088x^2 + 0.11972x^5 + 0.08550x^6 + 0.09816x^7 + 0.07194x^{16} + 0.34960x^{25}$<br>$\rho(x) = 0.58807x^{28} + 0.41193^{29}$ |
| 0.75 | 0.0392 | $\lambda(x) = 0.10805x + 0.09511x^2 + 0.01449x^3 + 0.13764x^4 + 0.10667x^5 + 0.05288x^6 + 0.01107x^{27} + 0.47408x^{30}$<br>$\rho(x) = 0.74161x^{24} + 0.25839^{25}$ |
| 0.70 | 0.0504 | $\lambda(x) = 0.05343x + 0.29406x^2 + 0.00896x^5 + 0.15571x^8 + 0.12189x^{11} + 0.19872x^{24} + 0.09572x^{45} + 0.02741x^{61}$<br>$+ 0.04056x^{64} + 0.00354x^{72}$<br>$\rho(x) = 0.76922x^{19} + 0.23077x^{20}$ |
| 0.65 | 0.0633 | $\lambda(x) = 0.10451x + 0.15652x^2 + 0.08057x^3 + 0.00056x^4 + 0.12151x^8 + 0.10485x^{12} + 0.10719x^{14} + 0.00771x^{20}$<br>$+ 0.31656x^{50}$<br>$\rho(x) = 0.000578x + 0.06089x^{14} + 0.47001x^{15} + 0.46852x^{20}$ |
| 0.60 | 0.0766 | $\lambda(x) = 0.11040x + 0.20804x^2 + 0.14163x^7 + 0.14858x^8 + 0.14438x^{25} + 0.08909x^{26} + 0.00748x^{45} + 0.15038x^{70}$<br>$\rho(x) = 0.00036x + 0.13063x^9 + 0.31068x^{12} + 0.49341x^{17} + 0.064915x^{18}$ |
| 0.55 | 0.0904 | $\lambda(x) = 0.16880x + 0.20994x^2 + 0.18095x^5 + 0.03846x^{14} + 0.02635x^{15} + 0.23454x^{17} + 0.05815x^{18} + .08280x^{30}$<br>$\rho(x) = 0.27631x^9 + 0.72369x^{10}$ |
| 0.50 | 0.1071 | $\lambda(x) = 0.14438x + 0.19026x^2 + 0.01836x^3 + 0.00233x^4 + 0.04697x^5 + 0.053943x^7 + 0.05590x^8 + 0.01290x^9$<br>$+ 0.00162x^{10} + 0.06159x^{13} + 0.13115x^{14} + 0.01481x^{16} + 0.00879x^{46} + 0.00650x^{48} + 0.00210x^{54} + 0.00099x^{55}$<br>$+ 0.11178x^{56} + 0.06238x^{57} + 0.05094x^{58} + 0.02230x^{65}$<br>$\rho(x) = 0.47575x^9 + 0.46847x^{11} + 0.02952x^{12} + 0.02626x^{13}$ |

few non zero values.

Their main advantage is that they can perform very close to Shannon limit, even with a suboptimal but fast, iterative decoding scheme.

In the case of reconciliation of binary strings, and hence for application to discrete-variable QKD, LDPC codes need to be specifically optimized for the BSC.

The LDPC code design optimization problem can be efficiently addressed with a genetic algorithm; *Differential Evolution* (DE) [15]. In particular, this solution was successfully applied for the BEC in [16] and for the BIAWGN channel in [17].

DE is an Evolutionary Optimization Algorithm, it maintains a population of $N$ $D$−dimensional vectors (code candidates) of real parameters respecting some constraints. This population evolves for a fixed number of generations or until a vector is found which meets a stopping criterion. The population is initialized to cover as much as possible of the parameter space. For each generation, DE mutates and recombines the current population to produce a trial population. Mutation is performed by adding the weighted difference of two population vectors to a third one.

Recombination is used to increase the diversity of the trial population: trial vectors are modified incorporating a small set of parameter values from a current population vector. A trial vector is incorporated into the current population if a cost function assigns to it a lower cost value than the cost value of the preceding vector, otherwise it is discarded.

LDPC codes can be represented as bipartite graphs [18]. One set of nodes, the check nodes, represents the set of parity-check equations which define the code; the other, the variable nodes, represents the elements of the codewords. A check (variable) node in the graph is called of degree $i$ if it is connected to $i$ variable (check) nodes. We denote by $\lambda_i$ ($\rho_i$) the fraction of edges which are connected to bit (check) nodes of degree $i$. Let $L$ be the maximum variable degree and $R$ the maximum check degree, we define an ensemble of LDPC codes by the generating functions $\lambda(x)$ and $\rho(x)$.

$$\lambda(x) := \sum_{i=2}^{L} \lambda_i x^{i-1} \qquad 0 \leq \lambda_i \leq 1$$
$$\rho(x) := \sum_{i=2}^{R} \rho_i x^{i-1} \qquad 0 \leq \rho_i \leq 1 \qquad (4)$$

We can express the code rate as a function of the coefficients of $\lambda(x)$ and $\rho(x)$:

$$\text{Rate} = 1 - \frac{\sum_{i=2}^{R} \rho_i/i}{\sum_{i=2}^{L} \lambda_i/i} \qquad (5)$$

The functions $\lambda(x)$ and $\rho(x)$ have $L + R - 2$ non zero coefficients. However not all these coefficients are independent: $\lambda(x)$ and $\rho(x)$ define degree distributions and must therefore be normalized, and we want all codes to be of the same rate in order to compare their thresholds.

In particular, to ensure that $\lambda(x)$ and $\rho(x)$ define a degree distribution we fix the coefficients corresponding to variable and check nodes of degree 2:

$$\lambda_2 = 1 - \sum_{i=3}^{L} \lambda_i, \qquad \rho_2 = 1 - \sum_{i=3}^{R} \rho_i \qquad (6)$$

We can set the code rate using a third coefficient, we use $\lambda_L$. From (5) and (6), one gets:

$$\lambda_L = \frac{\frac{1-\beta}{2} + \sum_{i=3}^{R} \rho_i(\frac{1}{i} - \frac{1}{2}) - \beta \sum_{i=3}^{L-1} \lambda_i(\frac{1}{i} - \frac{1}{2})}{\beta(\frac{1}{L} - \frac{1}{2})} \quad (7)$$

where $\beta = 1 - \text{Rate}$.

These three constraints leave a final number of $D = L + R - 5$ parameters each one associated with one of the non fixed coefficients of $\lambda(x)$ and $\rho(x)$. Finally we require the codes to be stable for crossover probabilities $p$ below their threshold, the stability condition for the BSC channel being given by [17]:

$$\lambda_2 \leq \frac{1}{2\sum_i (i-1)\rho_i \sqrt{p(1-p)}} \quad (8)$$

We have used discretized density evolution algorithm [19] to compute the cost function and evaluate the candidate codes. This algorithm calculates a threshold value for a random LDPC code with a fixed node and degree distribution. The threshold determines the limit of the error-free region asymptotically as the block length tends to infinity. Discretized density evolution guarantees that the predicted threshold is a lower bound of the real threshold.

The results we have obtained with this set of constraints are shown in Table I. For all rates the thresholds are very close to the Shannon limit. These thresholds are only achievable by infinite length codes, but experimental results obtained with finite length codes were not very different (see section IV-A). This is indeed not too surprising since the relevant length of codes we have used is quite large ($10^6$), adapted to the typical requirements of QKD where large blocks of data have to be processed together to minimize finite size effects [20].

## IV. EXPERIMENTAL RESULTS

In this section we discuss the experimental performances of Cascade and of our LDPC codes for block length of $10^6$. We have implemented Cascade as described in [5] and our LDPC codes are decoded with the belief propagation algorithm [21]. The remaining bit error probability is below $1.5 \cdot 10^{-6}$ and the remaining errors can be handled very efficiently by concatenation with a BCH code of very high rate (typically 0.998 [22]).

### A. Reconciliation Efficiency

As explained in Section I the performance of a reconciliation protocol can be evaluated by measuring the amount of information disclosed in this process. For the BSC with a crossover probability $p$, an ideal reconciliation protocol would reveal a fraction $h(p)$ while a real protocol reveals $f(p)h(p)$. We have represented the reconciliation efficiency $f(p)$ on Fig. 1 for Cascade and for our codes. The results that we have found with Cascade are very similar to those of Crepeau [5] or Brassard and Salvail [4]: Cascade performs well at low bit error rates where its efficiency differs only by $10\%$ from the Shannon limit of 1. However, its efficiency decreases gradually as the crossover probability increases.

A quick observation reveals that, in contrast with Cascade, the reconciliation efficiency $f(p)$ exhibits a saw behavior when our set of LDPC codes is used. The reason for this is that we have chosen a discrete number of codes. As each code has a certain threshold, a string with a measured error probability $p$ will be corrected with the code having the smallest threshold greater than $p$. The saw effect will be reduced as the number of LDPC codes used is increased. To illustrate this fact, we have also included in Fig.1 the smooth curve which would be the result of using an infinite number of LDPC codes.

As we can see on this figure, optimized LDPC codes can perform better than Cascade as soon as the error rate is above $2\%$. With our set of 9 LDPC codes, the performances are always better than Cascade when the error rate is above $5\%$. This gain of performance can significantly impact on the achievable secret key generation rate in practical QKD.

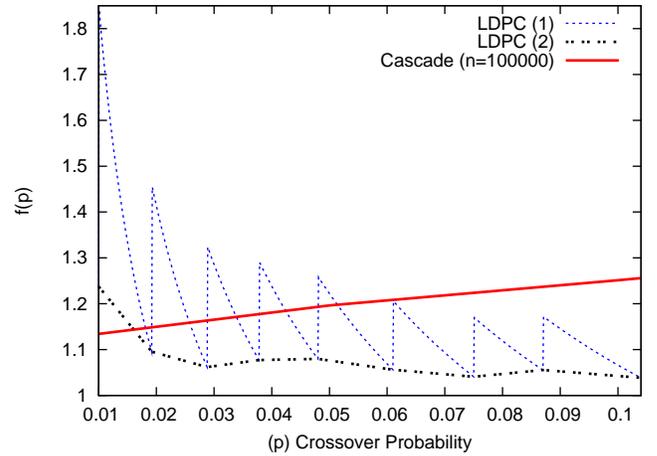

Fig. 1. Reconciliation Efficiency $f(p)$ achieved by LDPC codes and Cascade. (1) Using our set of 9 LDPC codes described in Table I (2) Extrapolated curve of f(p) for an infinite number of LDPC codes.

### B. Secret Key Rate and Local Randomization

As it appears from Eq. (3), the measure of the reconciliation efficiency $f(p)$ can be translated into a value of the achievable secret key rate $K_{real}(p)$, value that is indeed the true figure of merit for a practical QKD system.

In order to mitigate the saw effect produced by using a LDPC code non-adapted to the error-rate of the BSC, we have studied the impact of a possible improvement that can easily be implemented in practice: local randomization [23]. The idea is to make use of the LDPC codes in a error rate region close to their threshold, where their efficiency $f(p)$ is better. To achieve that in practice we use our freedom to worsen the error rate before performing the information reconciliation, by performing a local randomization on $X$, the string held by Alice. When the error rate $p$ is in the range $p \in (a, b)$ corresponding to the region of use of a given LDPC code of threshold $b$, we worsen the error rate from $p$ to $b$. To do that, Alice can flip each of her bits with a probability $e$. In order to have final error rate of $b$ we must take $e = \frac{b-p}{1-2p}$. From the point of view of Alice and Bob, the chain

$X$ is replaced by a chain $X'$ and the practical reconciliation reveals a fraction $fH(X'|Y) = f(b)h(b)$. From the point of view of Eve, the effect of local randomization is however worse: Eve holds $Z$, a noisy version of Alice's chain with $H(X|Z) = 1 - h(p) = h(q)$. Alice then flips each of her bits with probability $e$ and the effective error probability for Eve is $e(1-q) + (1-e)q = q + e - 2eq$. Applying Eq. (1) with $X'$ replacing $X$ leads to:

$$K_{lr}(p) = h(q + e - 2eq) - f(b)h(b) \qquad (9)$$

The comparison of what can be achieved using either Cascade or our LDPC codes is displayed on Fig. 2. The advantage of our reconciliation protocol can be well understood by considering the maximal admissible bit error rate. While it is less than 9.5 % with Cascade, it becomes very close the theoretical limit of 11 % with our protocol. This implies in practice that the maximum distance at which practical secure secret key distribution is possible will be extended when using our reconciliation protocol.

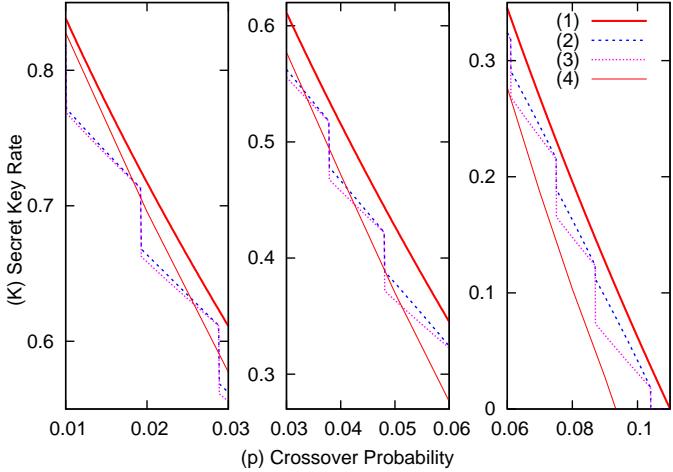

Fig. 2. Secret Key Rate for Cascade and LDPC codes. (1) Theoretical limit, (2) LDPC codes with local randomization , (3) LDPC codes without local randomization , (4) Cascade (n=100000)

## V. CONCLUSION

We have shown that LDPC codes can be used to reconcile two correlated discrete random variables. The results show that LDPC codes are a good alternative to Cascade. In terms of reconciliation efficacity they offer a similar behaviour for small crossover probabilities and a significant improvement for the crossover probabilities over 0.02. In terms of the interactivity LDPC codes need a single information exchange to reconcile the two variables while Cascade is very greedy in communication resources.

LDPC codes have been optimized for the BSC with thresholds near the channel capacity. This result can have a practical impact on the performance of QKD systems but also find a broader range of application to other scenarios where information-theoretic secret key agreement can be performed, such as the wiretap channel [24] or Maurer's satellite scenario [25].

ACKNOWLEDGMENT

The authors would like to thank Gilles Zémor for his comments and suggestions. We acknowledge support from the French Agence Nationale de la Recherche under project PROSPIQ (No. ANR- 06-NANO-041-05).